\documentclass[twoside]{article}
\usepackage{fleqn,espcrc2}

\usepackage{epsfig}
\usepackage[figuresright]{rotating}

\newcommand{\AmS}{{\protect\the\textfont2
  A\kern-.1667em\lower.5ex\hbox{M}\kern-.125emS}}

\title{Strange Pathways for Black Hole Formation}

\author{M. Prakash\address{Department of Physics and Astronomy, 
	State University of New York at Stony Brook,\\
	 Stony Brook,  New York 11794-3800, U.S.A}%
        \thanks{Research Supported by DOE Grant No. FG02-88ER-40388.}}

\begin{document}
\begin{abstract}
Immediately after they are born, neutron stars are characterized by an
entropy per baryon of order unity and by the presence of trapped
neutrinos. If the only hadrons in the star are nucleons, these effects
slightly reduce the maximum mass relative to cold, catalyzed matter.
However, if stangeness-bearing hyperons, a kaon condensate, or quarks
are also present, these effects result in an increase in the maximum
mass of up to $\sim 0.3{\rm M}_{\odot}$ compared to that of a cold,
neutrino-free star.  This makes a sufficiently massive proto-neutron
star metastable, so that after a delay of 10--100 seconds, the PNS
collapses into a black hole.  Such an event might be straightforward
to observe as an abrupt cessation of neutrinos when the instability is
triggered.
\vspace{1pc}
\end{abstract}
 
% typeset front matter (including abstract)
\maketitle
 
\def\be{\begin{eqnarray}}
\def\ee{\end{eqnarray}}

\section{INTRODUCTION}

Many fundamental and intrinsic properties of neutrinos play a crucial
role in astrophysical phenomena such as core-collapse supernovae.
Neutrinos drive supernova dynamics from beginning to end: they become
trapped within the star's core early in the collapse, forming a vast
energetic reservoir, and their eventual emission from the proto-neutron
star is prodiguous enough---containing nearly all the energy ($\sim  
10^{53}$ ergs) released
in the explosion---to dramatically control subsequent events.  

The large neutrino flux from a proto-neutron star (PNS) plays at least
two potentially important roles in the supernova environment.  First,
the supernova explosion itself may depend on neutrino heating to
propel a shock stalled by accretion. Second, the neutrino-driven wind
off the PNS that develops after shock lift-off may be a suitable site
for the synthesis of heavy elements produced by the rapid neutron
capture, or {\em r}-process, whose production site remains obscure.
Further progress in these issues will depend upon a thorough
understanding of the PNS neutrino emission.

The neutrinos of all flavors ($e, \mu, \tau$) emitted from newly
formed neutron stars in supernova explosions are the only direct probe
of the mechanism of supernovae and the structure of proto-neutron
stars.  The handful of neutrino events observed from SN 1987A
\cite{bion87,hira87} attest to  this fact. More intriguingly, the
three flavors of neutrino fluxes from a Galactic supernova could be
distinguished by the new generation of neutrino detectors. 
The supernova neutrino signals would furnish an
opportunity to probe the properties of neutrinos and dense matter in
regions that are inaccessible to terrestrial experiments.

\section{EVOLUTION OF PROTO-NEUTRON STARS} 

A proto-neutron star (PNS) forms in the aftermath of a successful
supernova explosion as the stellar remnant becomes gravitationally
decoupled from the expanding ejecta.  The essential microphysical
ingredients that govern the macrophysical evolution of the PNS in the
so-called Kelvin-Helmholtz epoch, during which the remnant changes
from a hot and lepton-rich PNS to a cold and deleptonized neutron
star, are the equation of state (EOS) of dense matter and its
associated neutrino opacity.  Among the characteristics of matter that
widely vary among EOS models are their relative compressibilities
(important in determining the theoretical neutron star maximum mass),
symmetry energies (important in determining the typical stellar radius
and in the relative $n, p, e$, and $\nu_e$ abundances) and specific
heats (important in determining the local temperatures).  These
characteristics play important roles in determining the matter's
composition, in particular the possible presence of strange components
(such as hyperons, a kaon condensate, or quark matter) \cite{prak97a}.
These characteristics also significantly affect calculated neutrino
opacities \cite{RPL98,Red99,BS98,BS99} and diffusion time scales.

The evolution of a PNS proceeds through several distinct stages
\cite{prak97a,burr86,burr90} and with various outcomes.  Immediately
following core bounce and the passage of a shock through the outer
PNS's mantle, the star contains an unshocked, low entropy core of mass
$M_c\simeq0.7$ M$_\odot$ in which neutrinos are trapped. The core is
surrounded by a low density, high entropy ($s=5-10$) mantle that is
both accreting matter falling through the shock and rapidly losing
energy due to beta decays and neutrino emission. The mantle extends to
the shock, which is temporarily stationary at a radius of about 200 km
prior to an eventual explosion.

After a few seconds, accretion becomes less important if the supernova
is successful and the shock lifts off the stellar envelope.  Extensive
neutrino losses and deleptonization of the mantle will have led to
loss of lepton pressure and collapse of the mantle.  If enough
accretion occurs, the star's mass could exceed the maximum mass of the
hot, lepton-rich matter and it collapses to form a black hole.  In
this event, neutrino emission is believed to quickly cease
\cite{burr88}.

Neutrino diffusion deleptonizes and heats the core on time scales of
10--15 s.  The core's maximum entropy is reached at the end of
deleptonization.  Strangeness could now appear, in which case the
maximum mass will decrease, leading to another possibility of black
hole formation \cite{TPL94,Bro94,BB94,PCL95,Gle95,KJ95,ELP96}.  If
strangeness does not appear, the maximum mass instead increases during
deleptonization and the appearance of a black hole would be less
likely.

The PNS is now lepton-poor, but it is still hot.  While the star has
zero net neutrino number, thermally produced neutrino pairs of all
flavors are abundant and dominate the emission.  The star cools, the
average neutrino energy decreases, and the neutrino mean free path
increases.  After approximately a few to 100 seconds (this time
depends on the form in which strangeness appears), the star becomes
transparent to neutrinos and finally achieves a cold, catalyzed
configuration.

Neutrino observations from a galactic supernova will illuminate these
stages.  The observables will concern time scales for deleptonization
and cooling and the star's binding energy.  Dimensionally, diffusion
time scales are proportional to $R^2(c\lambda)^{-1}$, where $R$ is the
star's radius and $\lambda$ is the effective neutrino mean free path.
This generic relation illustrates how both the EOS and the
composition, which determine both $R$ and $\lambda$, influence
evolutionary time scales.  Additional EOS dependence enters through
the rate at which the lepton number is lost from the star.  The total
binding energy is a stellar mass indicator.

\section{RESULTS OF SIMULATIONS}

\begin{figure}[ht]
\vspace{9pt}
%\framebox[55mm]{\rule[-21mm]{0mm}{43mm}}
\epsfig{file=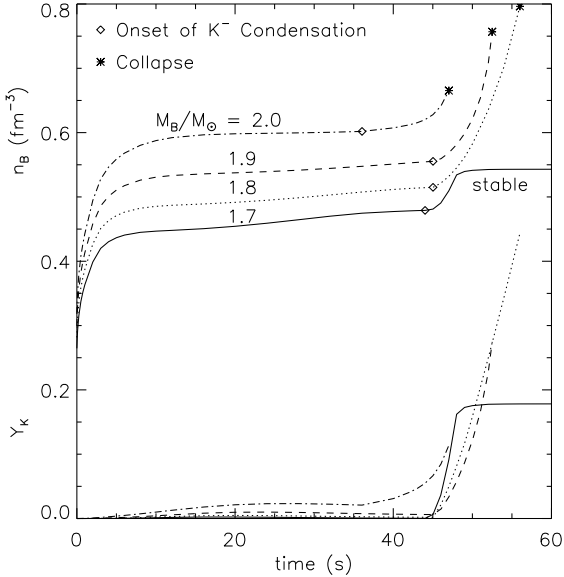, angle=0, height=3.in}
\vspace{-2pt}
\caption{Evolution of the central baryon density $n_B$ and the
kaon/baryon fraction $Y_K$ for stars with different baryonic masses.
Stars with masses larger $M_B=1.7 {\rm M}_\odot$ are
metastable. Diamonds indicate when kaon condensation occurs at the
stellar center, and asterisks denote when metastable stars become
gravitationally unstable.}
\label{fig:evol1}
\end{figure}

Detailed calculations of the evolution of PNSs, studying the
sensitivity of the results to the initial model, the total mass, the
underlying equation of state (EOS) and the possible presence of
hyperons and kaons, have been performed \cite{KJ95,Pon99}.  Pons, et
al. \cite{Pon99} show that the major effect on the neutrino signal
before the onset of any possible metastability is the PNS mass: larger
masses give rise to larger luminosities and generally higher average
emitted neutrino energies.  In addition, it was found that mass
windows for hyperonic metastable models could be as large as 0.3 M$_\odot$,
ranging from baryon masses $M_B=1.7$ M$_\odot$ to 2.0 M$_\odot$.  The
lifetimes of these stars decrease with stellar mass and
range from a few to longer than 100 s.  The detection of neutrinos
from SN 1987A over a timescale of 10-15 s is thus consistent with
either the formation of a stable PNS or a metastable PNS containing
hyperons, as long as its mass was less than about 0.1 M$_\odot$ below
the maximum mass for cold, catalyzed hyperonic matter.  Larger PNS
masses would lead to a collapse to a black hole on a timescale shorter
than that observed.

A similar situation could be encountered if the EOS allowed the
presence of other forms of ``exotic'' matter, manifested in the form
of a Bose condensate (of pions or kaons) or quarks
\cite{TPL94,PCL95,ELP96,SPL00}. Recently Pons, {\it et al.} \cite{Pon00b}
have studied the effect of kaon condensation on PNS evolution, by
employing an EOS which includes the effects of finite temperature and
neutrino trapping \cite{Pon00a}.  The phase transition from pure
nucleonic matter to the kaon condensate phase was described by means
of Gibbs' rules for phase equilibrium, which permit a mixed phase. In
the models explored there, the central densities were not large enough
to allow a pure condensed phase to exist.

Pons, {\it et al.} \cite{Pon00b} classify stars of different masses in three
main groups: i) stars in which the central density does not exceed the
critical value for kaon condensation, ii) stars that can form a mixed
phase core at the end of the Kelvin-Helmholtz epoch but remain stable,
and, iii) kaon condensed metastable stars that become unstable and
collapse to a black hole at the end of the Kelvin-Helmholtz epoch.
For stars in which the effects of kaon condensation are small (this
could be either because the star's mass is low enough to permit only a
very small region of the mixed phase or because condensation occurs as
a weak second order phase transition), the differences of the
predicted neutrino signal compared to PNSs composed of pure nucleonic
matter are very small.

In Fig. \ref{fig:evol1}, the evolution of stable (1.7 M$_\odot$) and
metastable stars are compared.  Both the central baryon number
densities and kaon fractions $Y_K$ are displayed.  In each case, the
time at which kaon condensation occurs is indicated by a diamond.
Asterisks mark the times at which the evolution of metastable stars
could not be further followed in our simulations, i.e., when a
configuration in hydrostatic equilibrium could not be found.  At this
time, the PNS is unstable to gravitational collapse into a black hole.
For the stable star, kaons appear after about 40 s.  Thereafter, the
star's central density increases in a short interval, about 5 s, until
a new stationary state with a mixed phase is reached.  The evolution
of the metastable stars is qualitatively different, inasmuch as the
central density increases monotonically from the time the condensate
appears to the time of gravitational collapse.
It is interesting that the lifetimes in all cases shown lie
in the narrow range 40--70 s (see Fig. \ref{fig:evol1}).  They decrease
mildly with increasing $M_B$. 

\begin{figure}[ht]
\vspace{9pt}
%\framebox[55mm]{\rule[-21mm]{0mm}{43mm}}
\epsfig{file=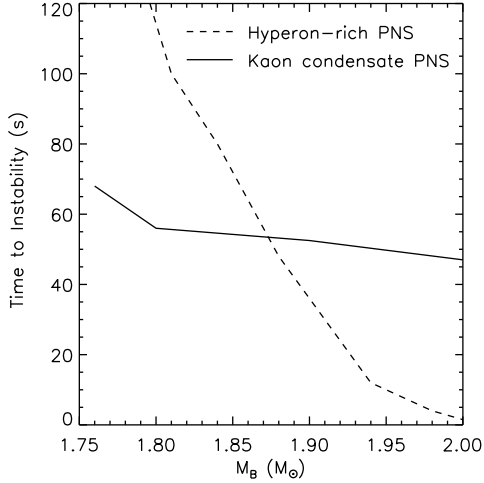, angle=0, height=2.5in}
\vspace{-2pt}
\caption{Lifetimes of metastable stars as a function
of the stellar baryon mass.  Solid lines show results for PNSs
containing kaon-condensates and dashed lines show the results of Pons, 
{\it et al.}  \cite {Pon99} for PNSs containing hyperons.}
\label{fig:ttc}
\end{figure}

In Fig. \ref{fig:ttc} the lifetimes versus $M_B$ for stars
containing hyperons ($npH$) and $npK$ stars are compared.  In both
cases, the larger the mass, the shorter the lifetime.  For kaon-rich
PNSs, however, the collapse is delayed until the final stage of the
Kelvin-Helmholtz epoch, while this is not necessarily the case for
hyperon-rich stars.

\section{SIGNALS IN DETECTORS}

\begin{figure}[hbt]
\vspace{9pt}
%\framebox[55mm]{\rule[-21mm]{0mm}{43mm}}
\epsfig{file=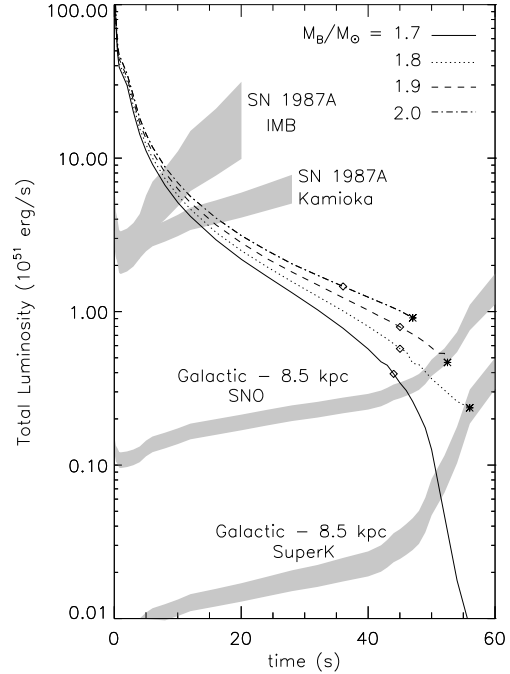, angle=0, height=3.5in}
\vspace{-2pt}
\caption{The evolution of the total neutrino luminosity for stars of
various baryon masses.  Shaded bands illustrate the limiting
luminosities corresponding to a count rate of 0.2 Hz in all detectors,
assuming a supernova distance of 50 kpc for IMB and Kamioka, and 8.5
kpc for SNO and SuperK. The width of the shaded regions represents
uncertainties in the average neutrino energy from the use of a
diffusion scheme for neutrino transport.}
\label{fig:lum1}
\end{figure}

In Fig. \ref{fig:lum1} the evolution of the total neutrino energy
luminosity is shown for different models. Notice that the drop in the
luminosity for the stable star (solid line), associated with the end
of the Kelvin-Helmholtz epoch, occurs at approximately the same time
as for the metastable stars with somewhat higher masses.  In all
cases, the total luminosity at the end of the simulations is below
$10^{51}$ erg/s.  The two upper shaded bands correspond to SN 1987A
detection limits with KII and IMB, and the lower bands correspond to
detection limits in SNO and SuperK for a future galactic supernova at
a distance of 8.5 kpc.  The times when these limits intersect the
model luminosities indicate the approximate times at which the count
rate drops below the background rate $(dN/dt)_{BG}=0.2$ Hz.

The poor statistics in the case of SN 1987A precluded a precise
estimate of the PNS mass.  Nevertheless, had a collapse to a black
hole occurred in this case, it must have happened after the detection
of neutrinos ended. Thus the SN 1987A signal is compatible with a late
kaonization-induced collapse, as well as a collapse due to
hyperonization or to the formation of a quark core.  More information
would be extracted from the detection of a galactic SN with the new
generation of neutrino detectors.

In SNO, about 400 counts are expected for electron antineutrinos from
a supernova located at 8.5 kpc.  The statistics would therefore be
improved significantly compared to the observations of SN 1987A.  A
sufficiently massive PNS with a kaon condensate becomes metastable,
and the neutrino signal terminates, before the signal decreases below
the assumed background.  In SuperK, however, up to 6000 events are
expected for the same conditions (because of the larger fiducial mass)
and the effects of metastability due to condensate formation in lower
mass stars would be observable.

\section{WHAT CAN WE LEARN?}

The calculations of Pons, {\it et al.} \cite{Pon00a} show that the variations
in the neutrino light curves caused by the appearance of a kaon
condensate in a stable star are small, and are apparently insensitive
to large variations in the opacities assumed for them.  Relative to a
star containing only nucleons, the expected signal differs by an
amount that is easily masked by an assumed PNS mass difference of
$0.01-0.02$ M$_\odot$.  This is in spite of the fact that, in some
cases, a first order phase transition appears at the star's center.
The manifestations of this phase transition are minimized because of
the long neutrino diffusion times in the star's core and the Gibbs'
character of the transition.  Both act in tandem to prevent either a
``core-quake'' or a secondary neutrino burst from occurring during the
Kelvin-Helmholtz epoch.

Observable signals of kaon condensation occur only in the case of
metastable stars that collapse to a black hole.  In this case, the
neutrino signal for a star closer than about 10 kpc is expected to
suddenly stop at a level well above that of the background in a
sufficiently massive detector with a low energy threshold such as
SuperK.  This is in contrast to the signal for a normal star of
similar mass for which the signal continues to fall until it is
obscured by the background.  The lifetime of kaon-condensed metastable
stars has a relatively small range, of order 50--70 s for the models
studied here, which is in sharp contrast to the case of hyperon-rich
metastable stars for which a significantly larger variation in the
lifetime (a few to over 100 s) was found.  This feature of kaon
condensation suggests that stars that destabilize rapidly cannot do so
because of kaons.

Pons, {\it et al.} \cite{Pon00a} determined the minimum lifetime for
metastable stars with kaons to be about 40 s by examining the most
favorable case for kaon condensation, which is obtained by maximizing
the magnitude of the optical potential.  The maximum optical potential
is limited by the binary pulsar mass constraint, which limits the
star's maximum gravitational mass to a minimum value of 1.44
M$_\odot$.  Therefore, should the neutrino signal from a future
supernova abruptly terminate sooner than 40 s after the birth of the
PNS, it would be more consistent with a hyperon- or quark-induced
instability than one due to kaon condensation.

It is important to note that the collapse to a black hole in the case
of kaon condensation is delayed until the final stages of the
Kelvin-Helmholtz epoch, due to the large neutrino diffusion time in
the inner core.  Consequently, to distinguish between stable and
metastable kaon-rich stars through observations of a cessation of a
neutrino signal from a galactic supernovae is only possible using
sufficiently massive neutrino detectors with low energy thresholds and
low backgrounds, such as the current SNO and SuperK, and future
planned (Super-DuperK) detectors.

\section{MATTER WITH QUARKS}

Strangeness appearing in the form of a mixed phase of strange quark
matter also leads to metastability.  Although quark matter is also
suppressed by trapped neutrinos \cite{PCL95,SPL00},
the transition to quark matter can occur at lower densities than 
the most optimistic kaon case, and the dependence of
the threshold density on $Y_L$ is less steep than that for kaons.  Thus, it
is an expectation that metastability due to the appearance of quarks,
as for the case of hyperons, might be able to occur relatively
quickly. 

\begin{figure}[ht]
\vspace{9pt}
%\framebox[55mm]{\rule[-21mm]{0mm}{43mm}}
\epsfig{file=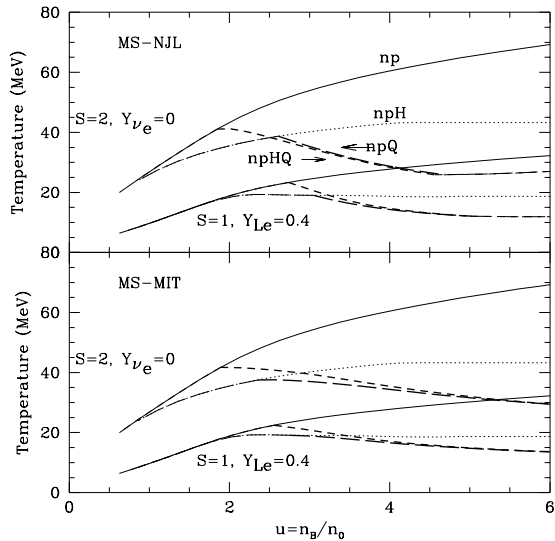, angle=0, height=3.in}
\vspace{-2pt}
\caption{Temperature versus density in units of $n_0$ for two PNS
evolutionary snapshots. The upper (lower) panel displays results for
the NJL (MIT bag) Lagrangian.  The parameters $\zeta=\xi=0$ in the
M\"uller-Serot (MS) hadronic Lagrangian are chosen.  Results are
compared for matter containing only nucleons (np), nucleons plus
hyperons (npH), nucleons plus quarks (npQ) and nucleons, hyperons and
quarks (npHQ).  Bold curves indicate the mixed phase region. See
Steiner, {\it et al.} \cite{SPL00} for more details.}
\label{fig:quarks}
\end{figure}

Steiner, {\it et al.} \cite{SPL00} have demonstrated that the temperature
along adiabats in the quark-hadron mixed phase is much smaller than
what is found for the kaon condensate-hadron mixed phase.  This could
lead to core temperatures which are significantly lower in stars
containing quarks than in those not containing quarks.
The temperature as a function of baryon density for fixed entropy and
net lepton concentration is presented in Fig.~\ref{fig:quarks},
which compares the cases ($s=1, Y_{L_e}=0.4$) and ($s=2, Y_{\nu_e}=0$)
both including and ignoring quarks.  The introduction of hyperons or
quarks lowers the Fermi energies of the nucleons and simultaneously
increases the specific heat of the matter, simply because there are
more components.  In the case of quarks, a further increase, which is
just as significant, occurs due to the fact that quarks are rather
more relativistic than hadrons.  The combined effects for quarks are
so large that, in some cases, an actual reduction of temperature with
increasing density occurs along an adiabat. This effect indicates that
the temperature will be smaller in a PNS containing quarks than in
stars without quarks.

The large reduction in temperature might also influence neutrino
opacities, which are generally proportional to $T^2$.  However, the
presence of droplet-like structures in the mixed phase
\cite{GS99,CGS00}, not considered here, will modify the specific heat.
In addition, these structures may dominate the opacity in the mixed
phase \cite{RBP00}.  However, a PNS simulation is necessary to
consistently evaluate the thermal evolution, since the smaller
pressure of quark-containing matter would tend to increase the star's
density and would oppose this effect.  Calculations of PNS evolution
with a mixed phase of quark matter, including the possible effects of
quark matter superfluidity \cite{CR00} are currently in progress and
will be reported separately.

\section {OUTLOOK} 

\subsection{ Neutrino Interferometry}

A titillating possibility would be a Hanbury-Brown Twiss
interferometric analysis using the neutrinos detected on earth to
determine the radial extent of the neutrinosphere.  For this purpose,
sufficient statistics and both accurate time and energy resolutions in
the detectors would be needed.

\subsection{$\pi^-$ and $K^-$
Dispersion Relations Through $\nu$-Nucleus Reactions}

The experimental program that would do the most to illuminate
theoretical issues permeating neutrino interactions in dense matter
would be studies of neutrino reactions on heavy nuclei, the only
direct way of probing the matrix elements of the axial current in
nuclear matter. Pioneering suggestions
in this regard have been put forth by Sawyer \& Soni \cite{sawy77,sawy78}, 
Ericson \cite{eric90}, and Sawyer \cite{sawy94}. The basic idea is to detect
positively charged leptons ($\mu^+~or~e^+$) produced in inclusive
experiments
\begin{eqnarray}
\bar \nu + X &\rightarrow & \mu^+~(\rm~{or}~e^+)~ + \pi^- 
          ~(\rm~{or}~K^-) + X 
\end{eqnarray}
which is kinematically made possible when the in-medium $\pi^-$ or
$K^-$ dispersion relation finds support in space-like regions.  The
sharp peaks at forward angles in the differential cross section versus
lepton momentum survive the 100-200 MeV width in the incoming GeV or
so neutrinos from accelerator experiments.  Calculations
of the background from quasi-elastic reactions
indicate that the signal would be easily detectable.

It is a pleasure to thank James Lattimer, Jose Pons,  and
Andrew Steiner, with whom the work reported here was performed.

\end{document}